\documentstyle[11pt,aaspp4,flushrt]{article}
\slugcomment{To appear in {\sl The Astrophysical Journal Letters}}

\lefthead{Shields et al.}
\righthead{NGC~4203 Nucleus}

\begin{document}

\title{Evidence for a Black Hole and Accretion Disk in the LINER NGC~4203$^1$}
\author{Joseph C. Shields$^2$, Hans-Walter Rix$^3$, Daniel H. McIntosh$^4$, 
Luis C. Ho$^5$, Greg Rudnick$^4$, Alexei~V.~Filippenko$^6$,
Wallace L. W. Sargent$^7$, and Marc Sarzi$^{3,8}$}
\altaffiltext{1} {Based on observations with the {\sl Hubble Space 
Telescope} obtained at STScI, which is operated by AURA, Inc., under NASA 
contract NAS5-26555.}
\altaffiltext{2}{Physics \& Astronomy Department, Ohio University, Athens, OH 45701; shields@phy.ohiou.edu}
\altaffiltext{3}{Max-Planck-Institut f{\"u}r Astronomie, K{\"o}nigstuhl 17,
Heidelberg, D-69117, Germany; rix@mpia-hd.mpg.de}
\altaffiltext{4}{Steward Observatory, University of Arizona, Tucson, AZ 85721;
dmac, grudnick@as.arizona.edu}
\altaffiltext{5}{The Observatories of the Carnegie Institution of Washington, 
813 Santa Barbara St., Pasadena, CA 91101-1292; lho@ociw.edu}
\altaffiltext{6}{Astronomy Department, University of California, Berkeley, CA
94720-3411; alex@astro.berkeley.edu}
\altaffiltext{7}{Palomar Observatory, Caltech 105-24, Pasadena, CA 91125;
wws@astro.caltech.edu}
\altaffiltext{8}{Dipartimento di Astronomina, Universit\`a di Padova, 
vicolo dell'Osservatorio 5 I-35122, Italy; sarzi@pd.astro.it}

\begin{abstract}

We present spectroscopic observations from the {\sl Hubble Space
Telescope} that reveal for the first time the presence of a broad
pedestal of Balmer-line emission in the LINER galaxy NGC 4203.  The
emission-line profile is suggestive of a relativistic accretion disk,
and is reminiscent of double-peaked transient Balmer emission observed
in a handful of other LINERs.  The very broad line emission thus constitutes
clear qualitative evidence for a black hole, and spatially resolved narrow-line
emission in NGC~4203 can be used to constrain its mass, $M_{BH }\le
6\times 10^6$ M$_\odot$ at 99.7\% confidence.  This value implies a
ratio of black-hole mass to bulge mass $\lesssim 7\times 10^{-4}$ in
NGC~4203, which is less by a factor of $\sim 3 - 9$ than the mean
ratio obtained for other galaxies.

The availability of an independent constraint on central black-hole
mass makes NGC~4203 an important testbed for probing the physics of
weak active galactic nuclei (AGNs).  Assuming $M_{BH }$ near the
detection limit, the ratio of observed luminosity to the Eddington
luminosity is $\sim 10^{-4}$.  This value is consistent with
advection-dominated accretion, and hence with scenarios in which an ion
torus irradiates an outer accretion disk that produces the observed
double-peaked line emission.  Follow-up observations will make it
possible to improve the black-hole mass estimate and study variability
in the nuclear emission.

\end{abstract}


\section{Introduction}
 
Spectroscopic surveys have revealed that a large fraction of nearby
early-type galaxies harbor low-ionization nuclear emission-line
regions, or LINERs (Heckman 1980\markcite{Heckman80}; Ho et
al. 1997a\markcite{Ho97a} and references therein).  The physical
understanding of these sources remains rudimentary, but there are
strong indications that at least some LINERs are weak versions of the
Seyfert or QSO phenomenon, and hence powered by accretion onto a
massive black hole.  LINERs are nonetheless distinct from classical
AGNs in terms of their characteristic (low) luminosity, emission-line
properties, and broad-band spectral energy distribution (Ho
1999\markcite{Ho99}).  These differences may follow from fundamental
disparities in the accretion process operative in luminous and weak
AGNs.

The study of emission-line behavior on small spatial scales within
galaxy nuclei provides one strategy for probing the energetics,
dynamics, and structure of LINERs and related objects.  Here we report
on observations acquired with the {\sl Hubble Space Telescope}\/ ({\sl HST}\/)
for the LINER NGC~4203.  The data provide dynamical constraints on a
black hole, and reveal line emission that may directly trace an accretion
flow in this source.  These observations and future follow-up studies
will provide an important framework for testing physical models for
the structure of LINERs, and the nature of black holes in galaxy
nuclei.

\section{Observations}

NGC~4203 was observed with the Space Telescope Imaging Spectrograph
(STIS) as part of a spectroscopic survey of nearby weakly active
galaxy nuclei.  The full details of this program will be reported
elsewhere (Rix et al., in preparation).  NGC~4203 is of S0
morphological type, seen nearly face-on, and its nucleus was
classified spectroscopically by Ho et al. (1997b)\markcite{Ho97b} as a
LINER 1.9 source.  NGC~4203 exhibits a recession velocity of 1088 km
s$^{-1}$, and we assume a distance for this source of 9.7 Mpc (Tully
1988\markcite{Tully88}).

{\sl HST} observations were acquired on 1999 April 18 UT, with the
$0\farcs2 \times 52\arcsec$ slit placed across the nucleus at position
angle (PA) = 105.6\arcdeg.  Two exposures totaling 1630 s and three
exposures totaling 2779 s were obtained with the G430L and G750M
gratings, resulting in spectra spanning 3300 -- 5700 \AA\ and 6300 --
6850 \AA\ with full width at half maximum (FWHM) spectral resolution
for extended sources of 10.9 and 2.2 \AA , respectively.  The
telescope was offset by 0\farcs05 ($\approx$ 1 pixel) along the slit
direction between repeated exposures.  The two-dimensional (2-D)
spectra were bias- and dark-subtracted, flat-fielded, shifted to a
common alignment, combined with repeated exposures to obtain single
blue and red spectra, cleaned of residual cosmic rays and hot pixels,
and corrected for geometrical distortion.  The data were wavelength-
and flux-calibrated via standard STSDAS procedures.

\section{Results}

\subsection{The Unresolved Nucleus}

Spectra of the nucleus of NGC~4203, obtained by coadding the 2-D
spectra over the central 0\farcs25 along the slit, are shown in Figure
1.  This extraction width represents approximately twice the FWHM for
the STIS point-spread function (PSF; i.e., FWHM = 0\farcs12).  The data
show strong, low-ionization forbidden line emission, consistent with
the classification of this source as a LINER.  The red spectrum
displays an additional striking feature in the form of a distinct
broad component of H$\alpha$, which contributes prominent
high-velocity shoulders to the line profile.  Ground-based
observations of this source obtained with the Palomar 5-m telescope in
1985 and reported by Ho et al. (1997c)\markcite{Ho97c} also show
evidence of broad H$\alpha$ emission, but in the earlier spectrum the
broad component is well represented by a Gaussian profile with a FWHM
of 1500 km s$^{-1}$, far less than the velocity difference of $\sim
7200$ km s$^{-1}$ between the profile shoulders evident in Figure
1{\sl a}.

We examined the properties of the broad H$\alpha$ emission in more
detail by removing the narrow contributions to the blended line.  A
synthetic profile for the narrow emission was constructed from the
[S~{\sc ii}] $\lambda\lambda$6716, 6731 lines, and used to model the
lines of [N~{\sc ii}] $\lambda\lambda$6548, 6583, and narrow
H$\alpha$.  The overall emission feature has a central core that
strongly resembles the broad H$\alpha$ line seen previously in the
Palomar spectrum, and we consequently included such a component of
``normal'' broad-line emission, matched to the profile parameters and
redshift of the ground-based observation, while fitting the overall
blend.  The [N~{\sc ii}] doublet was constrained to the flux ratio of
1:3.0 predicted by atomic parameters, and the line strengths were
otherwise adjusted in order to produce the smoothest residual profile.

Removal of the narrow lines and central peak of broad H$\alpha$
emission results in the H$\alpha$ line profile shown at the bottom of
Figure 1{\sl a}.  The profile of the remaining broad pedestal in the
6540 -- 6590 \AA\ region is sensitive to the details of the
decomposition; however, the profile remains distinctly double-peaked,
independent of the scaling for the subtracted components.  The broad
feature has a full-width near zero intensity of at least 12,500 km
s$^{-1}$, and a total flux of $\sim 1.2 \times 10^{-13}$ erg s$^{-1}$
cm$^{-2}$.  The line appears centered near the narrow-line redshift,
although the profile of the broad feature is highly asymmetric, with
stronger emission in the blue peak and a more extended wing on the red
side.  Similar emission is clearly visible in the profile of the
H$\beta$ feature (Fig. 1{\sl b}).

The scaling of the ``normal'' broad H$\alpha$ component employed in
the profile decomposition is 84\% that seen in the earlier Palomar
observation, which represents reasonable agreement when allowance is
made for photometric uncertainties in the ground-based data and
ambiguities in the fitting procedure.   The degree to which the broad
wings have varied is of considerable interest, given the variability
of double-peaked emission reported in some other LINERs (e.g.,
Storchi-Bergmann, Baldwin, \& Wilson 1993\markcite{Storchi93}; Halpern
\& Eracleous 1994\markcite{Halpern94}; Bower et al. 1996
\markcite{Bower96}).  To address this question, we conducted tests of
the observability of extended H$\alpha$ wings in the Palomar spectrum.
Shoulders on the H$\alpha$ profile resembling those in the {\sl HST}
data might have eluded detection in the Palomar spectrum if this
component was present in 1985, due to wavelength-dependent focus
variations and the strong stellar continuum measured through the
relatively large (2\arcsec $\times$ 4\arcsec) ground-based effective
aperture.

Since we cannot say for sure whether the broad shoulders on the Balmer
lines were present in 1985, it is unclear whether the appearance of
this emission in 1999 represents a transient event or a more stable
attribute of NGC~4203.  The fact that the central core of emission has
remained roughly constant suggests that the central source has not
changed appreciably, in which case the detection of the broad pedestal
of emission probably results primarily from the diminished
contamination by starlight in the {\sl HST} aperture, rather than
intrinsic variability.  {\sl HST} has revealed similar behavior in
NGC~4450, observed as part of our survey (Ho et
al. 2000a\markcite{Ho_et00a}).  The fact that {\sl two} sources out of a
relatively small sample of objects surveyed (24 galaxies total, of
which 8 are LINERs) were discovered to have double-peaked emission
lines suggests that small-aperture spectroscopy is an important
complement to variability studies for uncovering this phenomenon in
LINERs.

\subsection{Gas Kinematics and Mass Modeling}

The 2-D spectra of NGC~4203 exhibit spatially resolved narrow
emission, which is most readily apparent in H$\alpha$ and the [N~{\sc
ii}] lines.  These features are detected out to a distance of $\sim
1$\arcsec\ (47 pc) in both directions from the nucleus.  Line fluxes
and radial velocities were measured as a function of position along
the slit, by performing a simultaneous fit with $\chi^2$ minimization,
assuming Gaussian line profiles.  The [S~{\sc ii}] lines are visible
interior to $\pm$0\farcs5 from the nucleus, and are included in the
fits for that region.  The [O~{\sc i}] $\lambda\lambda$6300, 6364
lines are prominent in the nucleus but were not included in the fits;
emission in these lines is highly concentrated spatially, and the
[O~{\sc i}] profiles are also distinctly broader than those of the
other forbidden lines in the red spectrum (consistent with linewidth
vs. critical density correlations reported for other LINERs; e.g.,
Filippenko \& Halpern 1984\markcite{Filippenko84}).  The continuum
underlying the emission features was represented by a straight line,
and for columns near the nucleus the broad H$\alpha$ emission feature
was represented by the combination of three Gaussian profiles, which
was successful in isolating the narrow-line emission.

The radial velocity of the narrow-line gas is plotted as a function of
position in the top panel of Figure 2.  A distinct gradient is seen
through the central $\pm$0\farcs5, which is steepest in the
innermost $\pm$ 0\farcs1. Velocities level off and show evidence of a
significant decline in amplitude at larger distances, possibly due to
a warp in the gas disk.

The measurements illustrated in Figure 2 make it possible to probe the
central mass distribution in this galaxy, which is of particular
interest given the activity seen in the nucleus.  With only one slit
position, a number of assumptions are necessary to obtain a
well-constrained kinematic model. Here, we assumed (1) that the gas is
orbiting the nucleus in a coplanar disk at the local circular
velocity, (2) the {\it stellar} mass-to-light ratio $\Upsilon$ is
spatially constant and the stellar orbit distribution is approximately
isotropic, and (3) the stars dominate the total mass inside a
$1\farcs5\times 4\arcsec$ aperture (see Sarzi et al. 2000 for a
thorough description of the modeling).  We then used the acquisition
image taken with STIS to derive the deprojected, PSF-corrected light
distribution: it is well described by a cusp with $\rho_*(r)\propto
r^{-1.55}$ and a central, presumably nonstellar point source. Using
the isotropic, spherical Jeans equation, we can predict from
$\rho_*(r)$ the stellar velocity dispersion $\sigma_*$ over a
$1\farcs5 \times 4\arcsec$ aperture as a function of $\Upsilon$;
matching the observed $\sigma_*=124$~km s$^{-1}$ within this aperture
(Dalle Ore et al. 1991\markcite{Dalle91}) then dictates the value of
$\Upsilon$, to which we assign a 30\% error bar reflecting the
geometrical uncertainties in this derivation.  For any assumed
$M_{BH}$ one can then solve for the combination of disk inclination
and disk major axis that best matches the data.  The lines in the
bottom panel of Figure 2 represent such a model sequence in $M_{BH}$
where the disk orientation was chosen to optimize the match to the
data. While the ultra-broad H$\alpha$ emission implies the presence of
a black hole, the formal best fit to the extended rotation curve is
for a small black-hole mass (i.e., $\chi^2$ is minimized at $M_{BH} =
0$ M$_\odot$).  We thus obtain a formal upper limit of $M_{BH}\le
6\times 10^6$ M$_\odot$ at~99.7\%~confidence.

Examination of an archival WFPC2 $V$-band image reveals asymmetric
patchy absorption suggestive of an inclined ($\sim 50^\circ$) dusty
disk with a major-axis PA of $\sim$0\arcdeg.  If this orientation
applies to the nebular gas (as suggested by {\sl HST} imaging studies
of other LINERs; Pogge et al. 2000\markcite{Pogge00}), our slit PA is
$\sim 15\pm 20$\arcdeg\ from the disk minor axis.  Over this PA range
the projection factors change dramatically, and we cannot use the dust
lane information directly to constrain $M_{BH}$.  Formally, the
$\chi^2$ obtained with minor axis PA = 15\arcdeg\ is much worse than for
the best fit above.  We thus must await spectra at additional PAs or
offsets from the nucleus in order to obtain more direct constraints on
the disk orientation and black-hole mass (e.g., Bower et al. 
1998\markcite{Bower98}).

\section{Discussion}

\subsection{The Black Hole and its Galactic Host}

Studies of the velocity field for stars and occasionally gas in the
centers of nearby bulge-dominated galaxies have generated a list of
strong candidates for supermassive black holes (e.g., Kormendy \&
Richstone 1995\markcite{Kormendy95}; Ho 1998\markcite{Ho98}).  These
studies increasingly suggest that the black-hole mass is correlated
with the mass of the stellar bulge in the host galaxy.  We
consequently examined the black-hole mass in relation to the bulge
mass in the case of NGC~4203.  This galaxy has a total apparent
magnitude of $B_T = 11.61$, with $B-V = 0.97$ (de Vaucouleurs et
al. 1991\markcite{deVauc91}).  Burstein (1979)\markcite{Burstein79}
decomposed the $B$-band photometric profile of NGC~4203 into an
exponential disk and $r^{1/4}$-law bulge, and obtained a disk/bulge
luminosity ratio of 2.0.  For a representative mass-to-light ratio of
$\Upsilon_V \approx 6$, the corresponding bulge mass is $\sim 9 \times
10^9$ M$_\odot$; our upper limit of $M_{BH} \le 6 \times 10^6$
M$_\odot$ thus implies a black-hole/bulge mass ratio of $\lesssim7
\times 10^{-4}$.  Although this number is rather uncertain, it falls
near the low end of previous estimates for normal galaxies
($\sim$0.002 -- 0.006; Magorrian et al. 1998\markcite{Magorrian98};
van der Marel 1998\markcite{vdM98}) and overlaps with estimates
derived from reverberation measurements of Seyfert nuclei (Wandel
1999\markcite{Wandel99}).  A measurement or more stringent limit on
$M_{BH}$ in NGC~4203 would be of interest for gauging whether the
black-hole/bulge correlation applies in this source.

\subsection{The Black Hole and its AGN}

The luminous output of the LINER in NGC~4203 can be used as a
diagnostic of accretion onto the central black hole.  Ho et al. (2000b)
\markcite{Ho_et00b} have quantified the broad-band spectral energy 
distribution for this source, and estimate a total bolometric
luminosity for the nucleus of $9.5 \times 10^{40}$ erg s$^{-1}$.
For the limit on $M_{BH}$ obtained from~the nebular rotation curve,
this result implies a ratio to the Eddington luminosity of $L/L_{Edd}
\gtrsim 1 \times 10^{-4}$.

This result and the broad Balmer profiles in the NGC~4203 nucleus
have an appealing consistency in the framework of advection-dominated
accretion flow (ADAF) models for accretion onto the central black
hole.  ADAFs are expected to occur naturally in systems with
relatively low accretion rates (e.g., see Narayan, Mahadevan, \& 
Quataert 1998\markcite{Narayan98} for a review).  For an $\alpha$-disk
prescription for the accretion flow, an ion torus will form in the
inner disk, and advect a significant fraction of the accretion energy
into the black hole, when the accretion rate falls below a critical
value, $\dot M_{crit} \approx \alpha^2 \dot M_{Edd}$.  Here the
Eddington accretion rate is defined so that $L_{Edd} = 0.1\,\dot
M_{Edd}\,c^2$.  Adopting a plausible value of $\alpha \approx 0.3$
implies that $\dot M_{crit} \approx 0.1 \dot M_{Edd}$, corresponding
to $L \approx 0.1 L_{Edd}$, below which the luminosity is expected to
scale as $\dot M^2$.  The limiting value of $L/L_{Edd}$ in NGC~4203 
is in the advection-dominated regime, and corresponds to an
accretion rate $\dot M \approx 5 \times 10^{-4} M_\odot$ yr$^{-1}$;
an improved constraint on $M_{BH}$ is of obvious interest for testing
this physical picture.

ADAFs can account for many properties of luminous radio galaxies,
which often resemble NGC~4203 in displaying broad, double-peaked
Balmer emission lines (Eracleous \& Halpern 1994\markcite{Erac94}).
These profiles have been modeled in terms of a relativistic disk (Chen
\& Halpern 1989\markcite{Chen89}), with predicted properties notably
similar to those seen in NGC~4203 -- specifically, stronger emission
in the blue peak than in the red, and a sharper cutoff to the blue
shoulder than in the red.  The line emission presumably arises from an
outer part of the accretion flow that remains geometrically thin and
may be irradiated by the inner ion torus that is characteristic of
ADAFs.  Modeling of the emission profile can potentially yield an
estimate of the radius at which the transition from thin disk to ion
torus occurs, and we defer more detailed exploration of this
possibility to a later paper.

Alternative explanations exist for the double-peaked Balmer
lines.  Possibilities include emission from bipolar outflows or
anisotropically illuminated clouds, from gas surrounding binary black
holes, or from a tidally disrupted star or other asymmetric
circumnuclear structure (Eracleous et al. 1999\markcite{Erac99} and
references therein).  Studies of double-peaked emission profiles in
other LINERs have revealed evolution in the profiles over time, and
such observations can potentially be used to distinguish between these
possibilities, and to probe the structure of the accretion flow if a
disk is in fact responsible.  In the case of NGC~1097, variations in
the relative heights of the red and blue line peaks are suggestive of
emission from an orbiting ring that is azimuthally asymmetric in its
emissivity (Storchi-Bergmann et al. 1997\markcite{Storchi97}).  If the
elevated blue wing in NGC~4203 arises from an asymmetry of this type,
we might expect the peak in emission to eventually oscillate to the
red wing.  If the timescale for the asymmetry to propagate matches the
orbital timescale, then we can predict a rise in the red wing in
$\lesssim 4$ years, based on the velocity width of the emission-line
shoulders and our bound on the black-hole mass.  Periodic variability
on other (probably longer) timescales may be relevant if the asymmetry
propagates as a wave or via precession.  Follow-up observations thus
may yield further important constraints on the structure and evolution
of the accretion disk.

\section{Conclusions}

NGC~4203 is the latest addition to a small set of LINERs that exhibit
broad, double-shouldered Balmer lines.  The discovery of two such
objects in our {\sl HST} survey (NGC~4450 being the other; Ho et
al. 2000a\markcite{Ho_et00a}) suggests that such emission may be
common in LINERs, but often eludes detection in ground-based apertures
for which the contrast with the stellar continuum is weak.  The line
profile in NGC~4203 is noteworthy for its resemblance to emission
profiles seen in broad-line radio galaxies, which have been
interpreted as the signature of a thin outer accretion disk irradiated
by an inner ion torus, representing an advection-dominated flow onto a
black hole.

The nucleus of NGC~4203 also exhibits spatially resolved emission that
can be used to provide kinematic information on the underlying mass
distribution.  The existing observations at a single PA can be used to
restrict the underlying black-hole mass to $M_{BH} \le 6 \times 10^6$
M$_\odot$.  The availability of a mass estimate for the black hole is
of fundamental importance for studying the physics of the accretion
process.  Our limiting value for $M_{BH}$ is consistent
with a sub-Eddington accretion rate and formation of an ADAF, although a
more stringent limit could challenge this picture.  Follow-up
observations from space, or with small apertures under good seeing
conditions on the ground, will make it possible to improve the
estimate of the black-hole mass and study time evolution of 
the~broad~H$\alpha$~emission.

\acknowledgments

This research was supported financially through NASA grant NAG 5-3556,
and by GO-07361-96A, awarded by STScI, which is operated by AURA, Inc., for
NASA under contract NAS5-26555.  We thank T. Statler for helpful discussions,
and the referee, M. Eracleous, for valuable comments.


\clearpage
\begin{figure}
\plotone{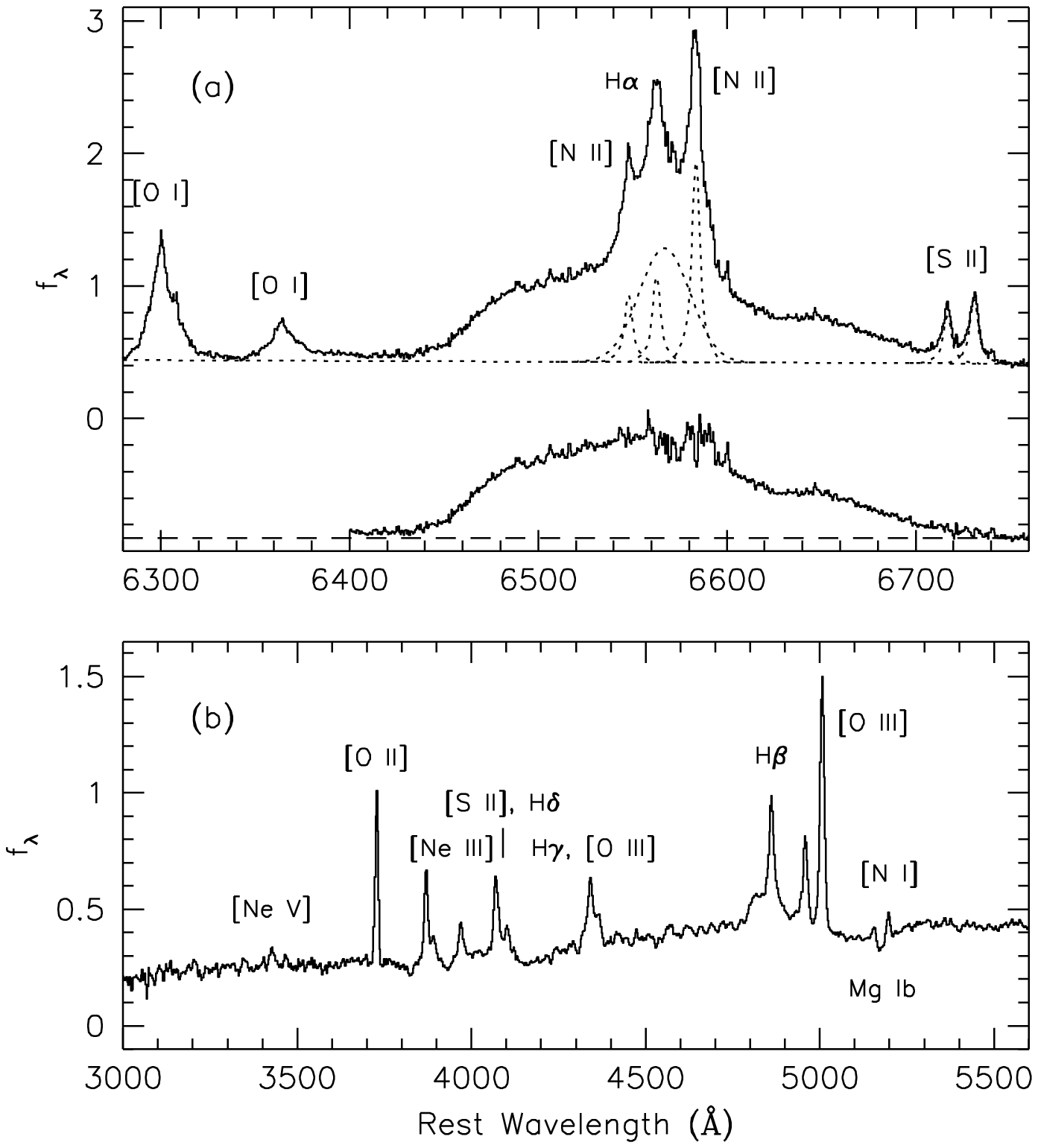}

\caption{STIS spectra of the nucleus of NGC~4203, acquired with the
({\sl a}) G750M and ({\sl b}) G430L gratings.  Flux densities are in
units of $10^{-15}$ erg s$^{-1}$ cm$^{-2}$ \AA $^{-1}$.  The dotted
curves in ({\sl a}) show model components adopted for the continuum,
narrow lines, and ``normal'' broad H$\alpha$ emission.  Subtraction of
these components yields the residual broad feature shown at the bottom
of the panel.}
\end{figure}
\begin{figure}
\plotone{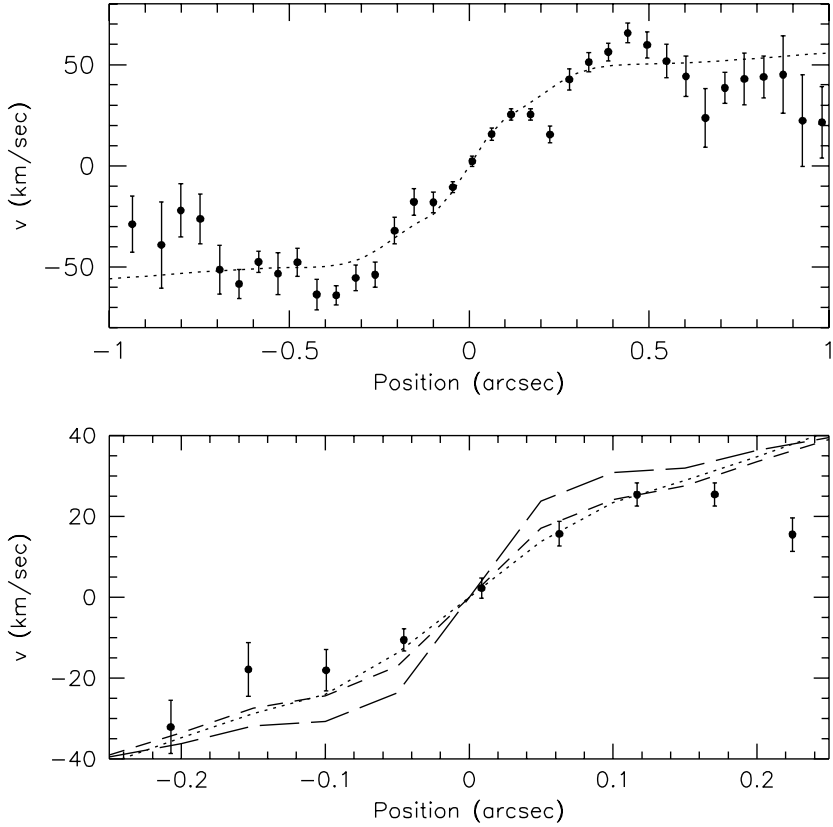}
\caption{{\sl Top panel:}  Emission-line gas velocity as a
function of position along the slit, measured relative to the nucleus.
Points represent the measured values, and the dotted line shows
the best-fitting model rotation curve with no black hole.
{\sl Bottom panel:} Expanded view of the central rotation curve.
The lines represent
predictions for the best-fitting model rotation curve with $M_{BH} =
0$ M$_\odot$ (dotted line), $4.0 \times 10^6$ M$_\odot$ (short-dashed
line), and $4.9 \times 10^7$ M$_\odot$ (long-dashed line).  In each
case, the model is constrained to reproduce the large-aperture stellar
velocity dispersion (see text for further details).  A $\chi^2$
analysis for a family of such fits implies that $M_{BH} \le 6 \times
10^6$ M$_\odot$ at 99.7\% confidence.}

\end{figure}

\end{document}